\documentclass[fleqn,10pt]{article}
\usepackage{fullpage}
\usepackage{amsfonts}
\usepackage{amsmath}
\usepackage{multirow}
\usepackage{longtable}
\usepackage{graphicx}
\usepackage{sidecap}
\usepackage{subcaption}
\usepackage[affil-it]{authblk}
\usepackage{hyperref}
\usepackage{cite}

\begin{document}

\title{\textbf{Correlations and dynamics of consumption patterns in social-economic networks}}
\date{\vspace{-5ex}}

\author[1]{Yannick Leo}
\affil[1]{\normalsize{Univ Lyon, ENS de Lyon, Inria, CNRS, UCB Lyon 1, LIP UMR 5668, IXXI, F-69342, Lyon, France}}

\author[1]{M\'arton Karsai\thanks{Corresponding author: marton.karsai@ens-lyon.fr}}

\author[2]{Carlos Sarraute}
\affil[1]{\normalsize{Grandata Labs, 550 15thStreet, San Francisco, California 94103, USA}}

\author[1]{Eric Fleury}

\maketitle

\begin{abstract}
We analyse a coupled dataset collecting the mobile phone communications and bank transactions history of a large number of individuals living in a Latin American country. After mapping the social structure and introducing indicators of socioeconomic status, demographic features, and purchasing habits of individuals we show that typical consumption patterns are strongly correlated with identified socioeconomic classes leading to patterns of stratification in the social structure. In addition we measure correlations between merchant categories and introduce a correlation network, which emerges with a meaningful community structure. We detect multivariate relations between merchant categories and show correlations in purchasing habits of individuals. Finally, by analysing individual consumption histories, we detect dynamical patterns in purchase behaviour and their correlations with the socioeconomic status, demographic characters and the egocentric social network of individuals. Our work provides novel and detailed insight into the relations between social and consuming behaviour with potential applications in resource allocation, marketing, and recommendation system design.
\end{abstract}

\section{Introduction}

Although consumption patterns are believed to be highly personal they still present certain similarities among people sharing some overall characteristics. Determinant factors are one's age, gender, or education level, while time and habitual environment can be also important. At the same time similarities may be induced via interaction on social ties, embedding an individual in a larger structure of a social network. \textit{Network effects} may further increase behavioural similarities as social influence arriving from connected neighbours could potentially bias one's purchasing preferences~\cite{Deaton1992Understanding,Bearden1989Measurement}. While these characters have been thoroughly studied in consumption behavioural research~\cite{Laroche2004Exploring,Loudon1993Consumer,Rodgers2003Gender,Sondhi2006Gender,Webster1997Resource} the effects of \textit{socioeconomic status} (SES) has been argued \cite{Dong2016Purchase,Pechey2015Behavior,Iqbal2011Buying} to play also an important role in determining consumption behaviour, i.e., the way for people to distribute their bounded financial capacities to purchase goods and services. The socioeconomic status of a person is determined by several intervening factors as income, educational level, ethnic, or occupation and its quantitative characterisation is a long lasting challenge. The uneven distribution of purchasing power among individuals goes hand in hand with the emergence and reservation of socioeconomic inequalities in general. Individual financial capacities restrict personal consumer behaviour, arguably correlate with one's purchasing preferences, and play indisputable roles in determining the socioeconomic position of an ego in the larger society~\cite{Deaton1992Understanding,Deaton1980Economics,Piketti2014Capital,Sernau2012Social,Hurst2015Social}. Moreover, socioeconomic status ~\cite{Bourdieu1984} plays an important role in shaping the global social network structure. Its entangled effect with status homophily~\cite{McPherson2001Birds,Lazarsfeld1954Friendship}, i.e., the tendency of people to connect to others at similar socioeconomic status, induces biases in tie creation preferences which lead to a stratified social structure~\cite{Sernau2012Social,Leo2016Socioeconomic,Grusky2011Theories} at large. This way people of the same social class may be better connected among each other, which further amplifies the emergence of common behavioural patterns characterising a given social group. The investigation of broader relations between the consumption patterns, socioeconomic status and demographic characters carries a great potential in understanding better rational social-economic behaviour~\cite{Fisher1987Fisher}. In addition, the dynamics of consumption patterns is a largely unexplored area. The dynamics of individual purchases may be strongly driven by periodic weekly fluctuations determined by ones occupation or demographic characters, which gives further motivations to study purchase behaviour from a time perspective.

To explore such problems Social Network Analysis (SNA) provides one promising direction~\cite{Wasserman1994Social}, due to its enormous benefit from the massive flow of human behavioural data provided by the digital data revolution~\cite{Lohr2012The,lazerscience2009,Abraham2010Computational}. On the other hand, although social behavioural data provides us detailed knowledge about the structure and dynamics of social interactions, it commonly fails to uncover the relationship between social and economic positions of individuals. Until now, the coupled investigation of individual social and economic status remained a great challenge due to lack of appropriate data recording such details simultaneously, even questions addressing correlation between consumption and social behaviour are at utmost interest.

In this paper we address these questions via the analysis of a dataset, which simultaneously records the mobile-phone communication, bank transaction history, and purchase sequences of millions of inhabitants of a Latin American country over several months. This corpus, one among the firsts at this scale and details, allows us to infer the socioeconomic status, consumption habits, and the underlying social structure of millions of connected individuals. Using this information our overall goal is to identify people with certain financial capacities, and to understand \textit{how much money they spend, on what they spend, when they spend, and whether they spend like their friends?} More precisely, we formulate our study around three research questions:
\begin{itemize}
\item Can one associate typical consumption patterns to people and to their peers belonging to the same or different socioeconomic classes, and if yes how much such patterns vary between individuals or different classes?
\item Can one draw relations between commonly purchased goods or services in order to understand better individual consumption behaviour?
\item Can one identify typical dynamical patterns of purchasing different goods and services by people with different gender, age, and socioeconomic status?
\end{itemize}
After reviewing the related literature in Section~\ref{sec:relatedwork}, we describe our dataset in Section~\ref{sec:data}, and introduce individual socioeconomic indicators to define socioeconomic classes in Section~\ref{sec:sociomeas}. In Section~\ref{sec:purchsocio} we show how typical consumption patterns vary among classes and we relate them to structural correlations in the social network. In Section~\ref{sec:purchnet} we draw a correlation network between consumption categories to detect patterns of commonly purchased goods and services. Subsequently in Section~\ref{sec:dynpatt} we address correlations between the weekly consumption dynamics of different purchasing categories with individual demographic and socioeconomic characters. Finally we present some concluding remarks and future research ideas.

\section{Related work}
\label{sec:relatedwork}

Earlier hypothesis on the relation between consumption patterns and socioeconomic inequalities, and their correlations with demographic features such as age, gender, or social status were drawn from specific sociological studies~\cite{chan2010social} and from cross-national social surveys~\cite{deaton1997analysis}. However, recently available large datasets help us to effectively validate and draw new hypotheses as population-large individual level observations and detailed analysis of human behavioural data became possible. These studies shown that personal social interactions, social influence \cite{Deaton1992Understanding,Bearden1989Measurement}, or homophily ~\cite{Wood2012Social} in terms of age or gender~\cite{Laroche2004Exploring,Loudon1993Consumer,Rodgers2003Gender,Sondhi2006Gender,Webster1997Resource,kovanen2013temporal} have strong effects on purchase behaviour, knowledge which led to the emergent domain of online social marketing~\cite{Felix2016Elements}. Yet it is challenging to measure correlations between social network, individual social status, and purchase patterns simultaneously~\cite{Dong2016Purchase,Pechey2015Behavior,Iqbal2011Buying}. One promising direction to map a society-large social network is provided by mobile phone data analysis \cite{Blondel2015A}. It has been shown that a social structure inferred from mobile-phone communications provides a $95 \%$ proxy of the original social network \cite{Eagle2009Inferring}. Even socioeconomic parameters can be estimated from communication networks~\cite{dong2014inferring,Specanovic2015Mobile,Blumenstock2010Mobile,Mao2015Quantifying,Blumenstock2015Predicting} or from external aggregate data~\cite{eagle2010network}. However, usually they do not come together with information on individual purchase behaviour, which can be the best estimated from anonymised purchased records \cite{Dong2016Purchase,Leo2016Socioeconomic}. In terms of spatial distribution, the relation between social networks and mobility patterns has been addressed in~\cite{Toole2015Coupling}, while the spatial variance of purchasing habits has also been investigated in~\cite{Sobolevsky2016Cities}. On the other hand, the dynamical variation of purchase patterns of different product categories is a largely unexplored area~\cite{Lareau2000Social} as it requires temporally detailed data recording individual purchase histories. In this paper we propose to explore these questions through the analysis of a combined dataset proposing simultaneous observations of social structure, economic status and purchase dynamics of millions of individuals.

Note that results presented below partially overlap with a related conference paper~\cite{Leo2016Correlations}, which has been published in the proceeding of the ASONAM'16 IEEE/ACM conference. The present paper extends this earlier work in several ways, including a detailed demographic analysis, deeper understanding on purchasing distributions in different categories, and the study addressing the dynamics of purchase patterns presented in Section~\ref{sec:dynpatt}.

\section{Data description}
\label{sec:data}

In the following we are going to introduce two datasets extracted from a corpus combining informations about the mobile phone interactions and purchase history of individuals.

\subsection*{DS1: Individual social-economic data with purchase distributions}
Communication data used in our study records the temporal sequence of 7,945,240,548 call and SMS interactions of 111,719,360 anonymised mobile phone users for $21$ months in a Latin American country. Each call detailed record (CDR) contains the time, unique caller and callee IDs, the direction (who initiate the call/SMS), and the duration of the interaction. At least one participant of each interaction is a client of a single mobile phone operator in the country, but other mobile phone users who are not clients of the actual provider also appear in the dataset with unique IDs. All unique IDs are anonymised as explained below, thus individual identification of any person is impossible from the data. Using this dataset we constructed a large social network where nodes are users (whether clients or not of the actual provider), while links are drawn between any two users if they interacted (via call or SMS) at least once during the observation period. We filtered out call services, companies, and other non-human actors from the social network by removing all nodes (and connected links) who appeared with either in-degree $k_{in}=0$ or out-degree $k_{out}=0$. We repeated this procedure recursively until we received a network where each user had $k_{in}, k_{out}>0$, i. e. made at least one out-going and received at least one in-coming communication event during the nearly two years of observation. After construction and filtering the network remained with 82,453,814 users connected by 1,002,833,289 links, which were considered to be undirected after this point.

To calculate individual economic estimators we used a dataset provided by a single Bank in the same country. This data records financial details of 6,002,192 people assigned with unique anonymised identifiers over $8$. The data provides time varying customer variables as the amount of their debit card purchases, their monthly loans, and static user attributes such as their billing postal code (zip code), their age and their gender.

A subset of IDs of the anonymised bank and mobile phone costumers were matched\footnote{Note, that the matching, data hashing, and anonymisation procedure was carried out through direct communication between the two providers (bank and mobile provider) without the involvement of the scientific partner. After this procedure only anonymised hashed IDs were shared disallowing the direct identification of individuals in any of the datasets.}. This way of combining the datasets allowed us to simultaneously observe the social structure and estimate economic status (for definition see Section~\ref{sec:sociomeas}) of the connected individuals. This combined dataset contained 999,456 IDs, which appeared in both corpuses. However, for the purpose of our study we considered only the largest connected component of this graph. This way we operate with a connected social graph of 992,538 people connected by 1,960,242 links, for all of them with communication events and detailed bank records available.

To study consumption behaviour we used purchase sequences recording the time, amount, merchant category code of each purchase event of each individual during the observation period of $8$ months. Purchase events are linked to one of the 281 merchant category codes (MCC) indicating the type of the actual purchase, like fast food restaurants, airlines, gas stations, etc. Due to the large number of categories in this case we decided to group MCCs by their types into $28$ purchase category groups (PCGs) using the categorisation proposed in~\cite{MCCAmExp}. After analysing each purchase groups $11$ of them appeared with extremely low activity representing less than 0.3\% (combined) of the total amount of purchases, thus we decided to remove them from our analysis and use only the remaining $K_{17}$ set of $17$ groups (for a complete list see Fig.\ref{fig:2}a). Note that the group named \emph{Service Providers} ($k_1$ with MCC $24$) plays a particular role as it corresponds to cash retrievals and money transfers and it represents around $70\%$ of the total amount of purchases. As this group dominates over other ones, and since we have no further information how the withdrawn cash was spent, we analyse this group $k_{1}$ separately from the other $K_{2\text{-}17}=K_{17}\backslash\{k_1\}$ set of groups.

This way we obtained DS1, which collects the social ties, economic status, and coarse grained purchase habit informations of $\sim 1$ million people connected together into a large social network. Note that although these people are connected into a single connected component, their observed social network is rather sparse as it is strongly limited by the intersecting user sets of the mobile provider and the bank. This way DS1 provides us meaningful informations on the dyadic and egocentric network level, however it does not captures many triads in the structure disallowing a study on the level of communities.

\subsection*{DS2: Detailed ego purchase distributions with age and gender}

From the same bank transaction trace of 6,002,192 users, we build a second data set DS2. This dataset collects data about the age and gender of individuals together with their purchase sequence recording the time, amount, and MCC of each debit card purchase of each ego. To receive a set of active users we extracted a corpus of 4,784,745 people that were active at least two months during the observation period. Then for each ego, we assigned a feature set $PV(u):\{ age_u, gender_u, SEG_u, r(c,u), w_u^{c}(d_i)\}$ where SEG assigns a socioeconomic group (for definition see Section~\ref{sec:sociomeas}), $r(c,u)$ is an ego purchase distribution vector defined as
\begin{equation}
r(c,u)=\frac{m_u^{c}}{\sum_{c} m_u^{c}}
\label{eq:rcu}
\end{equation}
that assigns the fraction of $m_u^{c}$ money spent by user $u$ on a merchant category $c$ during the observation period. We excluded purchases corresponding to cash retrievals and money transfers, which would dominate our measures otherwise. Finally, $w_u^{c}(d_i)$ is the ego weekly purchase distribution vector defined as the fraction of money spent by a user $u$ on a merchant category $c$ during the weekday $d_i$ such as $\sum_{i\in [0,6]}{w_u^{c}(d_i)}=1$. A minor fraction of purchases are not linked to valid MCCs, thus we excluded them from our calculations.

This way DS2 collects 3,680,652 individuals, without information about their underlying social network, but all assigned with a $PV(u)$ vector describing their personal demographic and purchasing features in details.

We introduced these two datasets separately because while in DS1 we have the advantage to access all informations, including purchase patterns, economic status, and social ties, its size is strongly limited by the intersection of the customer set of the mobile provider and the bank. On the other hand we will discuss several measures where social network information is not essential, thus we can exploit DS2, which is based on the considerably larger set of bank costumers even we have no information about their social ties.

\section{Measures of socioeconomic position}
\label{sec:sociomeas}

\begin{figure}[h!]
\centering
\includegraphics[width=1.0\textwidth]{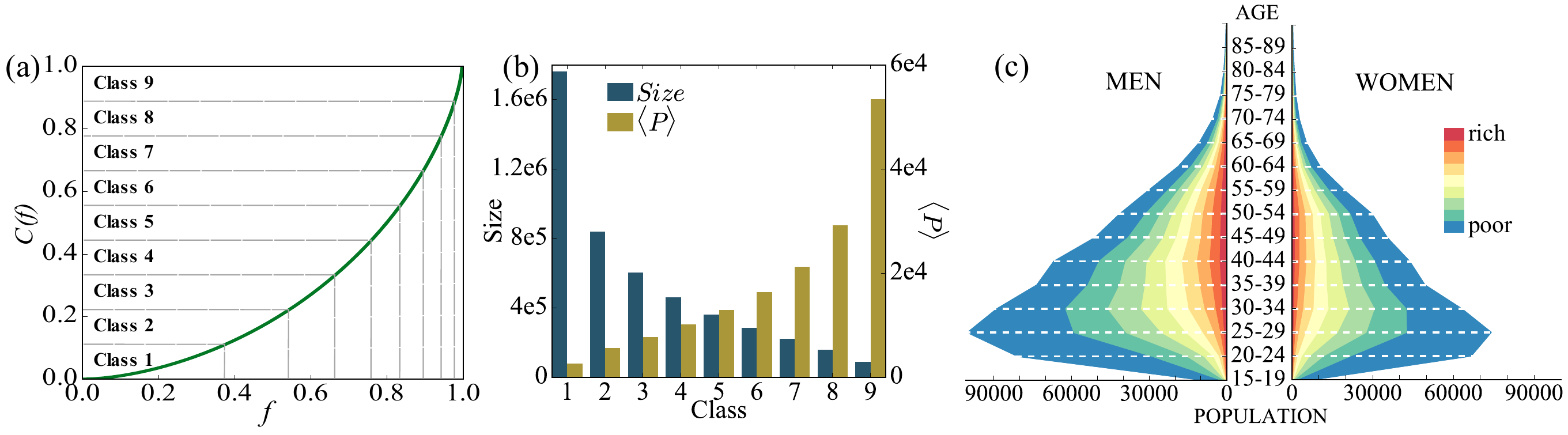}
\caption{\textbf{Social class and demographic characteristics} \textbf{(a)} Schematic demonstration of user partitions into 9 socioeconomic classes by using the cumulative AMP function $C(f)$. Fraction of egos belonging to a given class ($x$ axis) have the same sum of AMP $(\sum_u P_u)/n$ ($y$ axis) for each class. \textbf{(b)} Number of egos (green) and the average AMP $\langle P \rangle$ (in USD) per individual (yellow) in different classes. \textbf{(c)} Population pyramid of set of people in focus together with overall socioeconomic group informations depicted by colour codes (ranging from blue - poor to red - rich for the $9$ socioeconomic groups).
\label{fig:1}}
\end{figure}

To estimate the personal economic status we used a simple measure reflecting the consumption power of each individual. Starting from the raw data of DS2, which collects the amount and type of debit card purchases, we estimated the economic position of individuals as their average monthly purchase (AMP). More precisely, in case of an ego $u$ who spent $m_u(t)$ amount in month $t$ we calculated the AMP as
\begin{equation}
P_u=\frac{\sum_{t\in T}m_u(t)}{|T|_u}
\end{equation}
where $|T|_u$ corresponds to the number of active months of user $u$ (with at least one purchase in each month). After sorting people by their AMP values we computed the normalised cumulative distribution function of $P_u$ as
\begin{equation}
C(f)=\frac{\sum_{f'=0}^{f} P_u(f')}{\sum_{u} P_u}
\end{equation}
as a function of $f$ fraction of people. The $C(f)$ function in Fig.\ref{fig:1}a shows that AMP is distributed with a large variance signalling large economical imbalances just as suggested by the Pareto's law~\cite{Pareto1971Manual}. A conventional way to quantify the variation of this distribution is provided by the Gini coefficient $G$ \cite{Gastwirth1972The}, which characterises the deviation of the $C(f)$ function from a perfectly balanced situation, where wealth is evenly distributed among all individuals. In our case we found $G\approx 0.461$, which is relatively close to the World Bank reported value $G=0.481$ for the studied country~\cite{World2010Gini}, and corresponds to a Pareto index~\cite{Souma2000Physics} $\alpha=1.315$.

Subsequently we used the $C(f)$ function to assign egos into 9 economic classes (also called socioeconomic classes with smaller numbers assigning lower classes) such that the sum of AMP in each class $s_j$ was the same equal to $(\sum_u P_u)/n$ (Fig.\ref{fig:1}). We decided to use $9$ distinct classes based on the common three-stratum model~\cite{Brown2009Social}, which identifies three main social classes (lower, middle, and upper), and for each of them three sub-classes~\cite{Saunders1990Social}. There are several advantages of this classification: (a) it relies merely on individual economic estimators, $P_u$, (b) naturally partition egos into classes with decreasing sizes for richer groups and (c) increasing $\langle P \rangle$ average AMP values per egos (Fig.\ref{fig:1}b). Note that even the size of identified classes decreases with socioeconomic status we still have several thousands of individuals in the highest class $9$, which allows us to develop meaningful statistical claims about the behaviour of this set of people.

We also depict basic demographic informations about the different socioeconomic groups as a population pyramid. From Fig.\ref{fig:1}c we can conclude that the largest population is between age $25-29$, the largest fraction of people belong to the lowest socioeconomic group, and that there are more men than women in each age group.

\section{Socioeconomic correlations in purchasing patterns}
\label{sec:purchsocio}

\begin{figure*}[ht!]
\centering
\includegraphics[width=1.0\textwidth,angle=0]{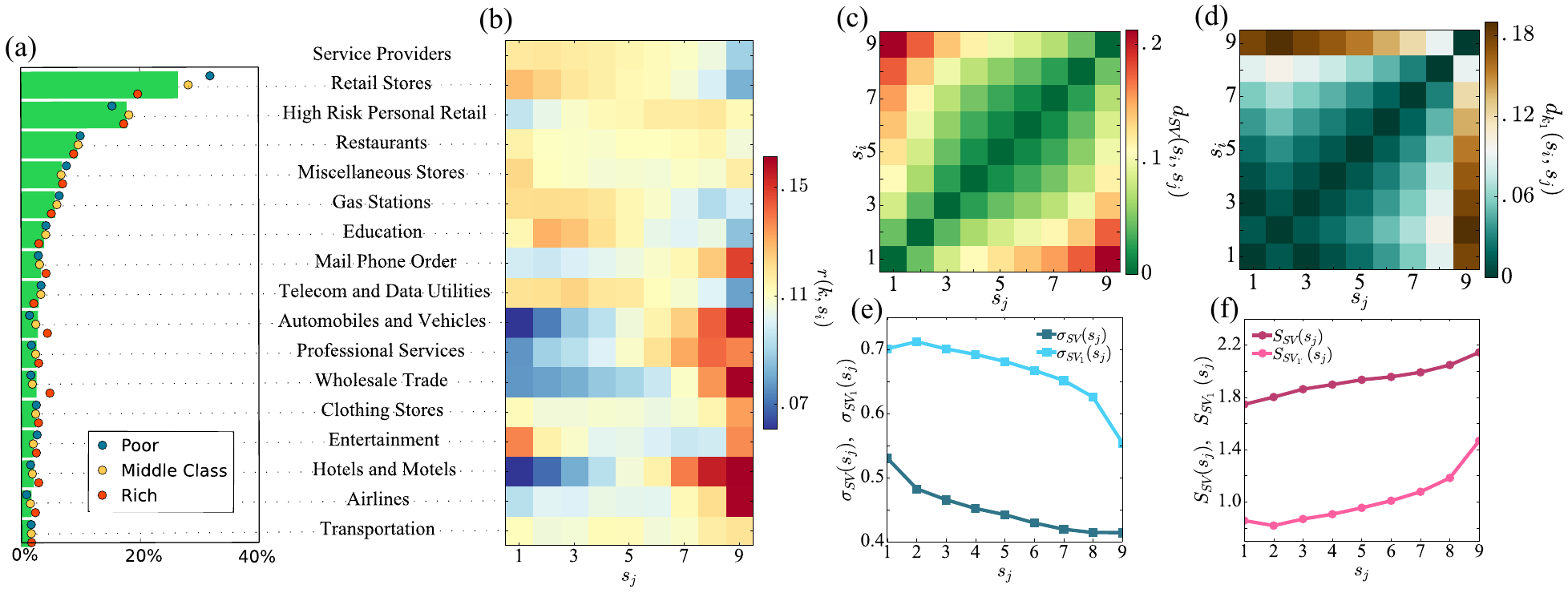}
\caption{\textbf{Consumption correlations in the socioeconomic network} \textbf{(a)} The histogram in green shows the distribution of the total amount of money spent on the PCGs $K_{2\text{-}17}$. Blue, yellow and red dots represents the values for the social class 1 (poorest), 5 (middle) and 9 (richest). \textbf{(b)} $r(k,s_i)$ distribution of spending in a given purchase category group $k\in K_{17}$ by different classes $s_j$. Distributions are normalised as in Eq.\ref{eq:rks}, i.e. sums up to $1$ for each category. \textbf{(c)} (resp. \textbf{(d)}) Heat-map matrix representation of $d_{SV}(s_i,s_j)$ (resp. $d_{k_{1}}(s_i,s_j)$) distances between the average spending vectors of pairs of socioeconomic classes considering PCGs in $K_{2\text{-}17}$ (resp. $k_1$). \textbf{(e)} Dispersion $\sigma_{SV}(s_j)$ for different socioeconomic classes considering PCGs in $K_{2\text{-}17}$ (dark blue) and the single category $k_1$ (light blue). The standard deviations of dispersion values are consistently around $STD_{SV} \approx 0.2$ ($STD_{SV_1} \approx 0.23$ for $k_1$) for each class. \textbf{(f)} Shannon entropy measures for different socioeconomic classes considering PCGs in $K_{2\text{-}17}$ (dark pink) and in $k_{1}$ (light pink).
\label{fig:2}}
\end{figure*}

In order to have an overall picture about possible differences between purchase preferences of people in different socioeconomic classes, we were looking for correlations between individuals in different socioeconomic classes in terms of their consumption behaviour on the level of purchase category groups. We analysed the purchasing behaviour of people in DS1 after categorising them into socioeconomic classes as explained in Section~\ref{sec:sociomeas}. In Fig.\ref{fig:2}a, without considering cash (\emph{Service Providers}) that represents around 68\% of the total spending, we show the percentage of total amount of money spent on each PCG. On average, bank clients spend 26.5\% of the total spending on \emph{Retail Stores}, 17.8\% on \emph{High Risk Personal Retail} and 9.5\% on \emph{Restaurants}, etc.. For each line of the histogram, percentages for the poorest, middle and richest social class are shown by coloured dots. It roughly points out major purchasing differences between social classes like in \emph{Retail Stores} the poorest people spend 31.8\% of their total amount of purchases (excluding cash retrieval) whereas the richest class spend only 19.6\% on the same category. 

To receive a finer information about the distributions of purchases, for each class $s_j$ we take every users $u\in s_j$ and calculate the $m_u^k$ total amount of purchases they spent on a purchase category group $k\in K_{17}$. Then we measure a fractional distribution of spending for each PCGs as:
\begin{equation}
r(k,s_j)=\frac{\sum_{u\in s_j} m^k_u}{\sum_{u\in s} m^k_u},
\label{eq:rks}
\end{equation}
where $s=\bigcup_{j}s_j$ assigns the complete set of users. In Fig.\ref{fig:2}b each line shows the $r(k,s_j)$ distributions for a PCG as the function of $s_j$ social classes, and lines are sorted (from top to bottom) by the total amount of money spent on the actual PCG\footnote{Note that in our social class definition the cumulative AMP is equal for each group and this way each group represents the same economic potential as a whole. Values shown in Fig.\ref{fig:2}b assign the total purchase of classes. Another strategy would be to calculate per capita measures, which in turn would be strongly dominated by values associated to the richest class, hiding any meaningful information about other classes.}. Interestingly, people from lower socioeconomic classes spend more on PCGs associated to essential needs, such as \emph{Retail Stores (St.)}, \emph{Gas Stations},  \emph{Service Providers} (cash) and \emph{Telecom}, while in the contrary, other categories associated to extra needs such as  \emph{High Risk Personal Retail} (Jewelry, Beauty), \emph{Mail Phone Order}, \emph{Automobiles}, \emph{Professional Services (Serv.)} (extra health services), \emph{Whole Trade} (auxiliary goods), \emph{Clothing St.}, \emph{Hotels} and \emph{Airlines} are dominated by people from higher socioeconomic classes. Also note that concerning \emph{Education} most of the money is spent by the lower middle classes, while  \emph{Miscellaneous St.} (gift, merchandise, pet St.) and more apparently \emph{Entertainment} are categories where the lowest and highest classes are spending the most.

\subsection{Purchase pattern similarities of socioeconomic classes}
\label{sec:purchsociosim}

From this first set of analysis we can already identify large differences in the spending behaviour of people from lower and upper classes. To further investigate these dissimilarities on the individual level, we consider the $K_{2\text{-}17}$ category set as defined in section~\ref{sec:data} (category $k_1$ excluded) and build a spending vector $SV(u)=[SV_2(u), ..., SV_{17}(u)]$ for each ego $u$. Here each item $SV_{k}(u)$ assigns the fraction of money $m_u^k/m_u$ what user $u$ spent on a category $k \in K_{2\text{-}17}$ out of his/her $m_u=\sum_{k \in K}{m_u^k}$ total amount of purchases. Using these individual spending vectors we calculate the average spending vector of a given socioeconomic class as $\overline{SV}(s_j) =\langle SV(u) \rangle_{u\in s_j}$. We associate $\overline{SV}(s_j)$ to a representative consumer of class $s_j$ and use this average vector to quantify differences between distinct socioeconomic classes as follows.

We measure the Euclidean metric between average spending vectors as:
\begin{equation}
d_{SV}(s_i,s_j) = \lVert \overline{SV}_k(s_i)-\overline{SV}_k(s_j)\rVert_2,
\end{equation}
where $ \lVert \vec{v}  \rVert_2=\sqrt{\sum_k v_k^2}$ assigns the $L^2$ norm of a vector $\vec{v}$ computed over $k \in K_{2\text{-}17}$ purchase categories. Note that the diagonal elements of $d_{SV}(s_i,s_i)$ are equal to zero by definition. However, in Fig.\ref{fig:2}c the off-diagonal green component around the diagonal indicates that the average spending behaviour of a given class is the most similar to neighbouring classes, while dissimilarities increase with the gap between socioeconomic classes. We repeated the same measurement separately for the single category of cash purchases (PCG $k_1$). In this case the Euclidean distance is defined between average scalar measures as $d_{k_1}(s_i,s_j)=\lVert \langle SV_1\rangle (s_i)-\langle SV_1\rangle(s_j) \rVert_2$. Interestingly, results shown in Fig.\ref{fig:2}d. indicates that here the richest social classes appear with a very different behaviour. This is due to their relative underspending in cash, which can be also concluded from Fig.\ref{fig:2}b (first row). On the other hand as going towards lower classes such differences decrease as cash usage starts to dominate.

To explain better the differences between socioeconomic classes in terms of purchasing patterns, we introduce two additional scalar measures. First, we introduce the dispersion of individual spending vectors as compared to their class average as
\begin{equation}
\sigma_{SV}(s_j) = \langle \lVert \overline{SV}_k(s_j)-SV_k(u) \rVert_2 \rangle_{u\in s_j},
\end{equation}
which appears with larger values if people in a given class allocate their spending very differently. Second, we also calculate the Shannon entropy of spending patterns as
\begin{equation}
S_{SV}(s_j) = \sum_{k\in K_{2\text{-} 17}}{-\overline{SV}_k(s_j) \log(\overline{SV}_k(s_j))}
\end{equation}
to quantify the variability of the average spending vector for each class. This measure is minimal if each ego of a class $s_j$ spends exclusively on the same single PCG, while it is maximal if they equally spend on each PCG. As it is shown in Fig.\ref{fig:2}e (light blue line with square symbols) dispersion decreases rapidly as going towards higher socioeconomic classes. This assigns that richer people tends to be more similar in terms of their purchase behaviour. On the other hand, surprisingly, in Fig.\ref{fig:2}f (dark pink line with square symbols) the increasing trend of the corresponding entropy measure suggests that even richer people behave more similar in terms of spending behaviour they used to allocate their purchases in more PCGs. These trends are consistent even in case of $k_1$ cash purchase category (see $\sigma_{SV_1}(s_j)$ function depicted with dark blue line in in Fig.\ref{fig:2}e) or once we include category $k_1$ into the entropy measure $S_{SV_{1}}(s_j)$ (shown in Fig.\ref{fig:2}f with light pink line).

\subsection{Purchase pattern similarities in the social network}
\label{sec:purchsociosimnet}

\begin{figure*}[ht!]
\centering
\includegraphics[width=1\textwidth,angle=0]{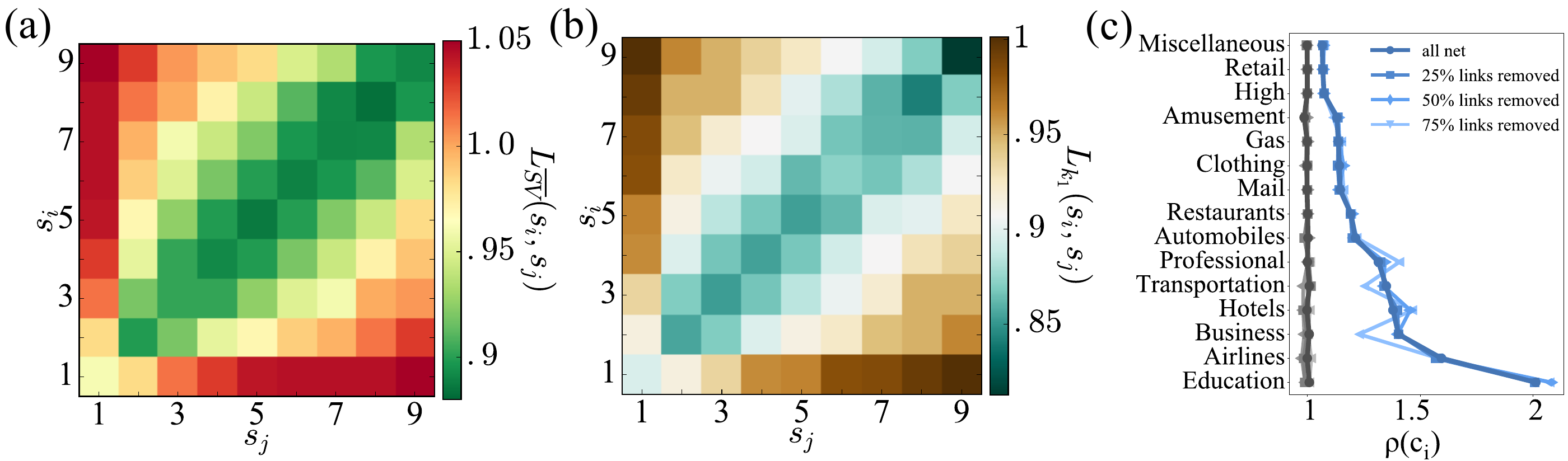}
\caption{\textbf{Network influence quantification.} \textbf{(a)} (resp. \textbf{(b)}) Heat-map matrix representation of the average $L_{SV}(s_i,s_j)$ (resp. $L_{k_1}(s_i,s_j)$) measure between pairs of socioeconomic classes considering PCGs in $K_{2\text{-}17}$ (resp. $k_1$). \text{(c)} Measure $\rho(c_{i})$ computed for real communication network (in blue) and for configuration model (in grey) for each PCG $k\in K_{2-17}$ sorted from smallest to greatest values of real network. Curves with different shades assign equivalent measures recalculated on the original network (all net) and jackknife samples~\cite{newman2003mixing} after the removal of the $25\%$, $50\%$, and $75\%$ of links.
\label{fig:3}}
\end{figure*}

To complete our investigation we characterise the effects of social relationships on the purchase habits of individuals. We address this problem through an overall measure quantifying differences between individual purchase vectors of connected egos positioned in the same or different socioeconomic classes. More precisely, we consider each social tie $(u,v)\in E$ connecting individuals $u\in s_i$ and $v\in s_j$, and for each purchase category $k$ we calculate the average absolute difference of their purchase vector items as
\begin{equation}
d^k(s_i,s_j)=\langle | SV_k(u)-SV_k(v)|\rangle_{u\in s_i, v\in s_j}.
\end{equation}
Following that, as a reference system we generated a corresponding configuration model network~\cite{Newman2010Networks} by taking randomly selected edge pairs from the underlying social structure and swapped them without allowing multiple links and self loops. In order to vanish any residual correlations we repeated this procedure in $5\times |E|$ times. This randomisation keeps the degree, individual economic estimators $P_u$, the purchase vector $SV(u)$, and the assigned class of each people unchanged, but destroys any structural correlations between egos in the social network, consequently between socioeconomic classes as well. After generating this reference structure we computed an equivalent measure $d_{rn}^k(s_i,s_j)$ but now using links $(u,v)\in E_{rn}$ of the randomised network. We repeated this procedure $100$ times and calculated an average $\langle d_{rn}^k\rangle (s_i,s_j)$. In order to quantify the effect of the social network we simply take the ratio
\begin{equation}
L_k(s_i,s_j)=\frac{d^k(s_i,s_j)}{\langle d_{rn}^k\rangle (s_i,s_j)}
\end{equation}
and calculate its average $L_{SV}(s_i,s_j)=\langle L_k(s_i,s_j)\rangle_k$ over each category group $k\in K_{2\text{-}17}$ or respectively $k_1$. This measure shows whether connected people have more similar purchasing patterns than one would expect by chance without considering any effect of homophily, social influence or structural correlations. Results depicted in Fig.\ref{fig:3}a for $L_{SV}(s_i,s_j)$ appear with a strong diagonal component, which indicate that the purchasing patterns of individuals connected in the original structure are actually more similar than expected from the random reference structure. On the other hand people from remote socioeconomic classes appear to be less similar than one would expect from the uncorrelated case (indicated by the $L_{SV}(s_i,s_j)>1$ values typical for upper classes in Fig.\ref{fig:3}a). Note that we found the same correlation trends in cash purchase patterns by measuring $L_{k_1}(s_i,s_j)$ as shown in Fig.\ref{fig:3}b. These observations do not clearly assign whether homophily~\cite{McPherson2001Birds,Lazarsfeld1954Friendship} or social influence~\cite{Deaton1992Understanding} induce the observed similarities in purchasing habits but undoubtedly clarifies that social ties (i.e. the neighbours of an ego) and socioeconomic status play deterministic roles in the emerging similarities in consumption behaviour.

These results show that some categories appear to be more similar when considering social ties, thus they might be more sensitive to inter-personal influence than others. In order to quantify directly inter-personal similarities in purchase habits we take the network $G=(V,E)$ and for each PCG $c_i$ we introduce a measure $\rho$ defined as follow :
\begin{equation}
\rho(c_{i},E)=\Big \langle \frac{r(c_{i},u)}{\langle r(c_{i},u) \rangle} \times \frac{r(c_{i},v)}{\langle r(c_{i},v) \rangle} \Big \rangle_{(u,v)\in E}
\label{eq:corMccSelf}
\end{equation}
Here $\rho(c_{i})$ quantifies the tendency that two connected egos $(u,v)$ spend commonly in a same category $c_i$. If their behaviour is independent than $ \sum_{(u,v)\in E}{r(c_{i},u) r(c_{i},v)} = (\sum_{(u,v)\in E}{r(c_{i},u)}) (\sum_{(u,v)\in E}{r(c_{i},v)})$ and so $\rho(c_{i})=1$. On the other hand, if their spending patterns on a given category show similarities, the dependence measure is $\rho(c_{i})>1$. In Fig.\ref{fig:3}c, given the communication network, the values of $\rho(c_{i})$ are all greater than $1$, meaning that connected people spend in any PCGs in a similar way thus their behaviour is not independent, i.e. assortative patterns characterise the network. This is even more evident once we compare these results to equivalent ones, measured on the configuration model graph. There $\rho$ is always close to $1$ as ties were randomised thus connected neighbours are selected independently by definition. A conventional way to test the variance of such assortative mixing patterns is by computing, for each group, the standard deviation of $\rho(c_{i})$ using the jackknife method~\cite{newman2003mixing}. In our case, due to the several purchase category groups considered, and the large number of links in the network, the iterative edge removal process to compute the standard deviation in this way~\cite{newman2003mixing} is computationally not feasible. Nevertheless, in order to demonstrate the variance of the observed assortative mixing patterns, we removed the $25\%$, $50\%$, and $75\%$ of links randomly and re-calculated the $\rho(c_{i})$ values for each category using the remaining links. As shown in Fig.3(c), the recomputed curves vary weakly as we increase the fraction of removed links, this way suggesting a small variance and strong robustness of the assortative spending patterns in the social network. These results also demonstrate the dependences between the diversity of similarity of different PCGs. Some purchase category groups like \emph{Education}, \emph{Airlines}, \emph{Business Services} or \emph{Hotels} are closely dependent over ties ($\rho \approx 2$), while ties have not much impact on purchases associated to everyday necessities like daily supermarket spendings.

\section{Purchase category correlations}
\label{sec:purchnet}

To study consumption patterns of single purchase categories PCGs provides a too coarse grained level of description. Hence, to address our second question we use DS2 and we downscale from the category group level to the level of single merchant categories. We are dealing with 271 categories after excluding some with less than 100 purchases and the categories linked to money transfer and cash retrieval (for a complete list of IDs and name of the purchase categories considered see Table~\ref{table:mcc}). As in Section~\ref{sec:data} we assign to each ego $u$ a personal vector $PV(u)$ of four socioeconomic features: the age, the gender, the social economic group, and the distribution $r(c_i,u)$ of purchases in different merchant categories made by the central ego. Our aim here is to obtain an overall picture of the consumption structure at the level of merchant categories and to understand precisely how personal and socioeconomic features correlate with the spending behaviour of individuals and with the overall consumption structure. 

\begin{figure*}[ht!]
\centering
\includegraphics[width=1.0\textwidth,angle=0]{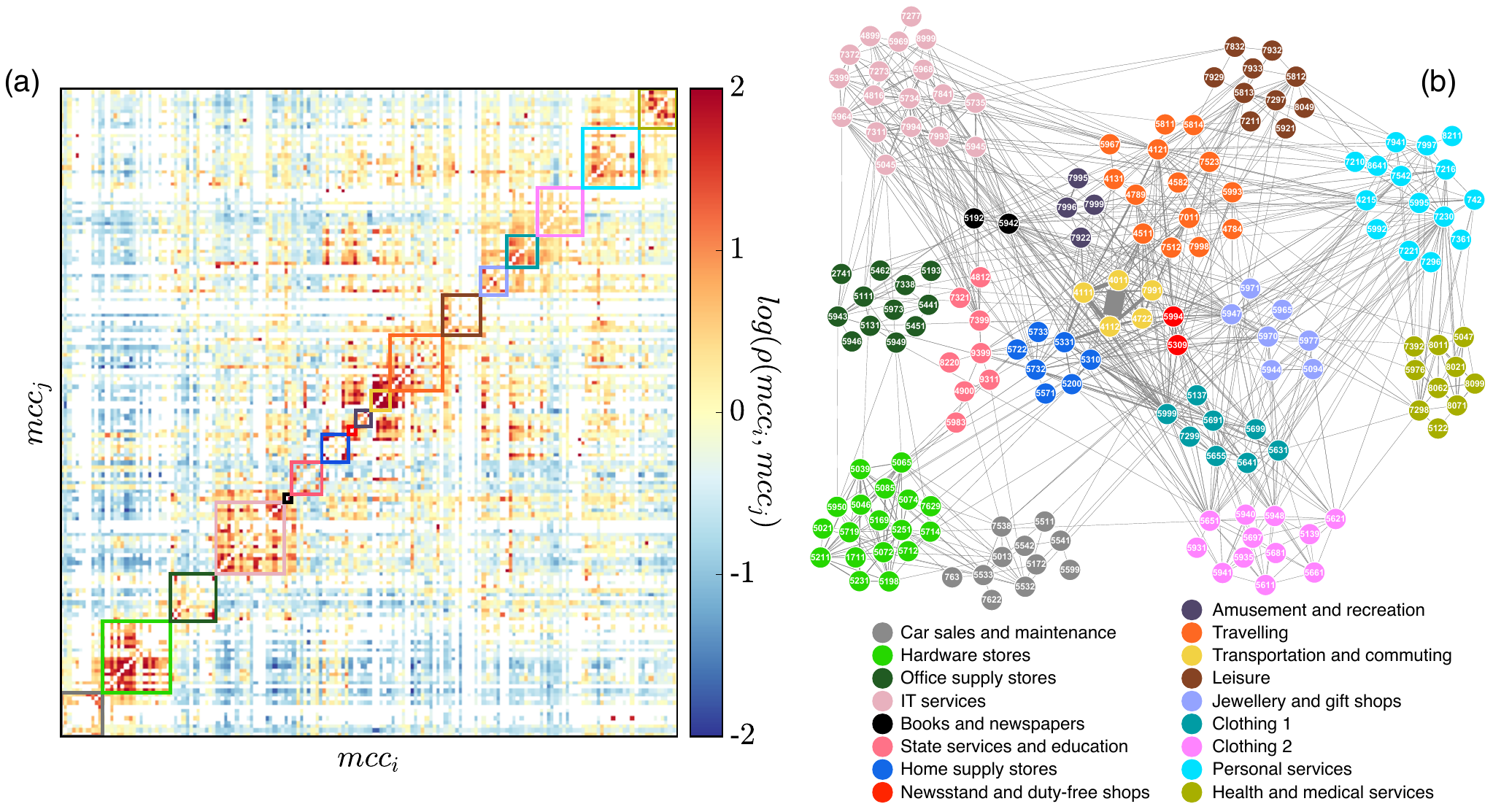}
\caption{\textbf{Merchant category correlation matrix and graph} \textbf{(a)} 163$\times$163 matrix heat-map plot corresponding to $\rho(c_{i}, c_{j})$ correlation values (see Eq.~\ref{eq:corMcc}) between categories. Colours scale with the logarithm of correlation values. Positive (resp. negative) correlations are assigned by red (resp. blue) colours. Diagonal components represent communities with frames coloured accordingly.\textbf{(b)} Weighted $G_{\rho}^>$ correlation graph with nodes  annotated with MCCs (see Table~\ref{table:mcc}). Colours assign 17 communities of merchant categories with representative names summarised in the figure legend.}
\label{fig:4}
\end{figure*}

As we noted in Eq.~\ref{eq:rcu}, the purchase spending vector $r(c_{i},u)$ of an ego quantifies the fraction of money spent on a category $c_i$. Using the spending vectors of $n$ number of individuals we define an overall measure between categories as
\begin{equation}
\rho(c_{i}, c_{j})= \Big \langle \frac{r(c_{i},u)}{\langle r(c_{i},u) \rangle} \times \frac{r(c_{j},u)}{\langle r(c_{j},u) \rangle} \Big \rangle_{u \in V} 
\label{eq:corMcc}
\end{equation}
This symmetric formulae quantifies how much people spend on a category $c_{i}$ if they spend on an other $c_{j}$ category or vice versa. Therefore, if $\rho(c_{i}, c_{j})>1$, the categories $c_{i}$ and $c_{j}$ are positively correlated and if $\rho(c_{i}, c_{j})<1$, categories are negatively correlated. Using $\rho(c_{i}, c_{j})$ we can define a weighted correlation graph $G_{\rho}=(V_{\rho},E_{\rho},\rho)$ between categories $c_i\in V_{\rho}$, where links $(c_{i}, c_{j})\in E_{\rho}$ are weighted by the $\rho(c_{i}, c_{j})$ correlation values. The weighted adjacency matrix of $G_{\rho}$ is shown in Fig.\ref{fig:4}a as a heat-map matrix with logarithmically scaling colours. Importantly, this matrix emerges with several block diagonal components suggesting present communities of strongly correlated categories in the graph.

To identify categories which were commonly purchased together we consider only links with positive correlations. Furthermore, to avoid false positive correlations, we consider a $10\%$ error on $r$. The worst global overestimation is reached when the two numerators $r(c_{i},.)$ are overestimated and the two denominators $\langle r(c_{i},.) \rangle$ are underestimated by $10\%$. In this case, $\rho_{approx.} = \frac{110\%^2}{90\%^2}\rho_{real} = 1.49\rho_{real}$. We insure to obtain only true positive correlations by taking $\rho\geq 1.5$ values. In addition, to consider only representative correlations we take into account category pairs which were commonly purchased by at least $1000$ consumers. This way we receive a $G_{\rho}^>$ weighted sub-graph of $G_{\rho}$, shown in Fig.\ref{fig:4}b, with 163 nodes and 1664 edges with weights $\rho(c_{i}, c_{j})>1.5$.

To identify communities in $G_{\rho}^>$ indicated by the correlation matrix in Fig.\ref{fig:4}a we applied a graph partitioning method based on the Louvain algorithm~\cite{blondel2008fast}. We obtained 17 communities depicted with different colours in Fig.\ref{fig:4}b and as corresponding coloured frames in Fig.\ref{fig:4}a. Interestingly, each of these communities group a homogeneous set of merchant categories, which could be assigned to similar types of purchasing activities (see legend of Fig.\ref{fig:4}b). In addition, this graph indicates how different communities are connected together. Some of them, like \textit{Transportation, IT} or \textit{Personal Serv.} playing a central role as connected to many other communities, while other components like \textit{Car sales and maintenance} and \textit{Hardware St.}, or \textit{Personal} and \textit{Health and medical Serv.} are more connected among each other than to the rest of the network. Some groups emerge as standalone communities like \textit{Office Supp. St.}, while others like \textit{Books and newspapers} or \textit{Newsstands and duty-free shops} appear as bridges despite their small sizes.

Note that the main categories corresponding to everyday necessities, related to food (\emph{Supermarkets}, \emph{Food St.}) and telecommunication (\emph{Telecommunication Serv.}), do not appear in this graph. Since they are responsible for the majority of total spending, they are purchased necessarily by everyone without obviously enhancing the purchase in other categories, thus they do not appear with strong correlations. This observation may confirm results obtained in Section~\ref{sec:purchsocio} where PCGs such as \emph{Miscellaneous Stores} and \emph{Retail Stores} were not correlated with the social ties.

Next we turn to study possible correlations between purchase categories and personal features. First we assign an average feature set $AFS(c_i)=\{\langle age(c_i) \rangle, \langle gender(c_i) \rangle, \langle SEG(c_i) \rangle \}$ to each of the 271 categories. The average $\langle v(c_i) \rangle$ of a feature $v\in \{ age, gender, SEG\}$ assigns a weighted average value computed as:
\begin{equation}
\langle v(c_i) \rangle = \frac{\sum_{u \in \{ u\}_i}\alpha_i(v_u)v_u}{\sum_{u \in \{ u\}_i}\alpha_i(v_u)},
\label{eq:vc}
\end{equation}
where $v_u$ denotes a feature of a user $u$ from the $\{ u\}_i$ set of individuals who spent on category $c_i$. Here
\begin{equation}
\alpha_i(v_u)=\sum_{(u \in \{ u\}_i | v_u=v)} \frac{r(c_i,u)}{n_i(v_u)}
\end{equation}
corresponds to the average spending on category $c_i$ of the set of users from $\{ u\}_i$ sharing the same value of the feature $v$ and $n_i(v_u)$ denotes the number of such users. In other words, e.g. in case of $v=age$ and $c_{742}$, $\langle age(c_{742}) \rangle$ assigns the average age of people spent on Veterinary Services ($MCC=742$) weighted by the amount they spent on it. In case of $v=gender$ we assigned $0$ to females and $1$ to males, thus the average gender of a category can take any real value between $[0, 1]$, indicating more females if $\langle gender(c_i) \rangle\leq 0.5$ or more males otherwise.

\begin{figure*}[ht!]
\centering
\includegraphics[width=1.0\textwidth,angle=0]{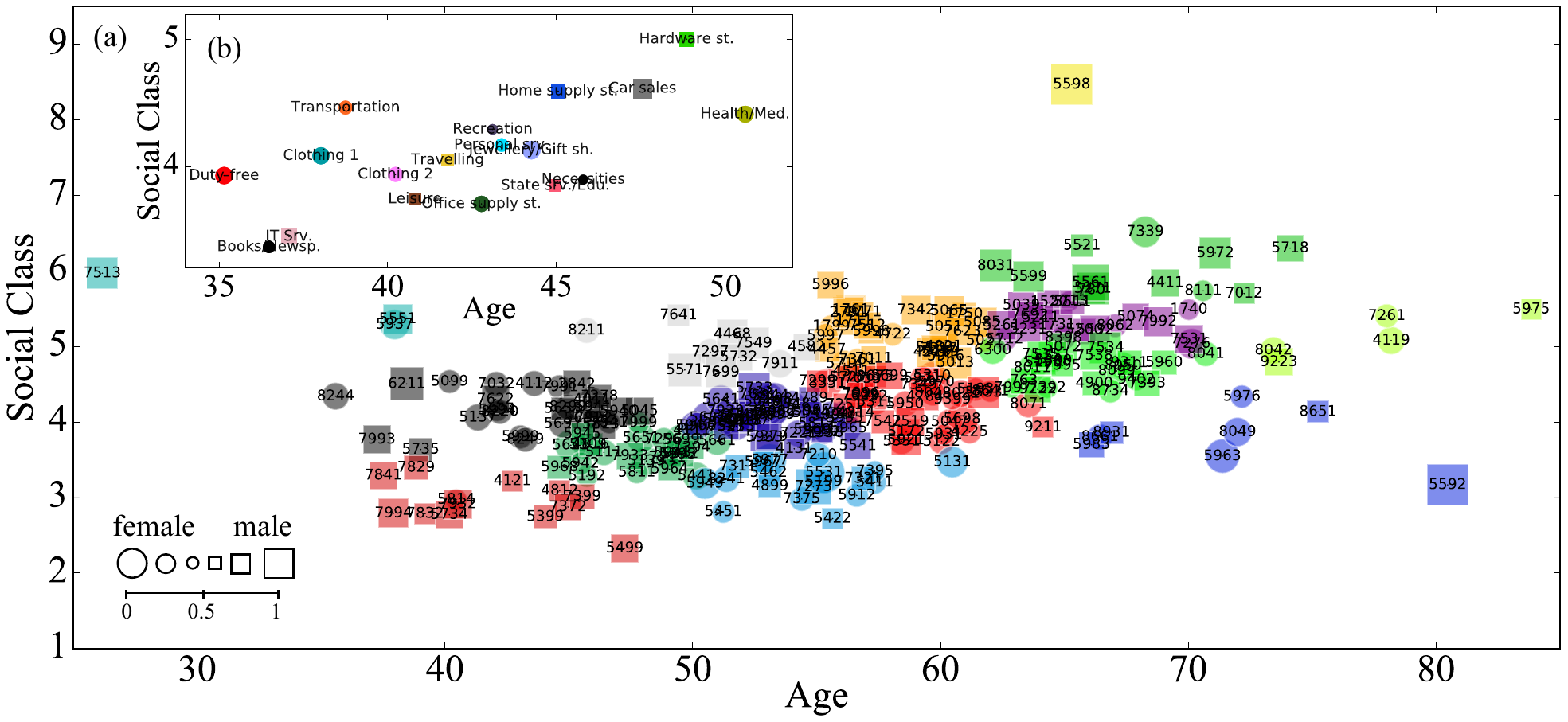}
\caption{\textbf{Socioeconomic parameters of merchant categories} \textbf{(a)} Scatter plot of $AFS(c_i)$ triplets (for definition see Eq.~\ref{eq:vc} and text) for 271 merchant categories summarised in Table~\ref{table:mcc}. Axes assign average age and SEG of purchase categories, while gender information are assigned by symbols. The shape of symbols assigns the dominant gender (circle-female, square-male) and their size scales with average values. \textbf{(b)} Similar scatter plot computed for communities presented in Fig.\ref{fig:3}b. Labels and colours are explained in the legend of Fig.\ref{fig:3}a.\label{fig:5}}
\end{figure*}

We visualise this multi-modal data in Fig.\ref{fig:5}a as a scatter plot, where axes scale with average age and SEG, while the shape and size of symbols correspond to the average gender of each category. To further identify correlations we applied k-means clustering~\cite{Bishop1995Neural} using the $AFS(c_i)$ of each category. The ideal number of clusters was $15$ according to several criteria: Davies-Bouldin Criterion, Calinski-Harabasz criterion (variance ratio criterion) and the Gap method \cite{Tibshirani2001Estimating}. Colours in Fig.\ref{fig:5}a assign the identified k-mean clusters.

The first thing to remark in Fig.\ref{fig:5}a is that the average age and the SEG assigned to merchant categories are positively correlated with a Pearson correlation coefficient $0.42$ ($p<0.01$). In other words, elderly people used to purchase from more expensive categories, or alternatively, wealthier people tend to be older, in accordance with our intuition. At the same time, some signs of gender imbalances can be also concluded from this plot. Wealthier people commonly appear to be males rather than females. A Pearson correlation measure between gender and SEG, which appears with a coefficient $0.29$ ($p<0.01)$ confirms it. On the other hand, no strong correlation was observed between age and gender from this analysis. Note that these correlations are not only significant but they are in line with our earlier observations on demographics depicted in Fig.~\ref{fig:1}.

To have an intuitive insight about the distribution of merchant categories, we take a closer look at specific category codes (summarised in Table~\ref{table:mcc}). As seen in Fig.\ref{fig:5}a elderly people tend to purchase in specific categories such as \emph{Medical Serv.}, \emph{Funeral Serv.}, \emph{Religious Organisations}, \emph{Motorhomes Dealers}, \emph{Donation}, \emph{Legal Serv.}. Whereas categories such as \emph{Fast Foods}, \emph{Video Game Arcades}, \emph{Cinema}, \emph{Record St.}, \emph{Educational Serv.}, \emph{Uniforms Clothing}, \emph{Passenger Railways}, \emph{Colleges-Universities} are associated to younger individuals on average. At the same time, wealthier people purchase more in categories as \emph{Snowmobile Dealers}, \emph{Secretarial Serv.}, \emph{Swimming Pools Sales}, \emph{Car Dealers Sales}, while poorer people tend to purchase more in categories related to everyday necessities like \emph{Food St.}, \emph{General Merch.}, \emph{Dairy Products St.}, \emph{Fast Foods} and \emph{Phone St.}, or to entertainment as \emph{Billiard} or \emph{Video Game Arcades}. Typical purchase categories are also strongly correlated with gender as categories more associated to females are like \emph{Beauty Sh.}, \emph{Cosmetic St.}, \emph{Health and Beauty Spas}, \emph{Women Clothing St.} and \emph{Child Care Serv.}, while others are preferred by males like \emph{Motor Homes Dealers}, \emph{Snowmobile Dealers}, \emph{Dating/Escort Serv.}, \emph{Osteopaths}, \emph{Instruments St.}, \emph{Electrical St.}, \emph{Alcohol St.} and \emph{Video Game Arcades}.

Finally we repeated a similar analysis on communities found in Fig.\ref{fig:4}b, but computing the $AFS$ on a set of categories that belong to the same community. Results in Fig.\ref{fig:5}b disclose positive age-SEG correlations as observed in Fig.\ref{fig:5}a, together with somewhat intuitive distribution of the communities.

\section{Correlations in purchase dynamics}
\label{sec:dynpatt}

The daily and weekly rhythms of people may be influenced by several external factors such as occupation, family status, habitual place, hobbies, just to mention a few. In addition personal variables, like age and gender, may also play a role here and contribute to individual circadian patterns, which were found recently \cite{Aledavood2015Daily}. Although all these parameters may influence the purchase dynamics of individuals~\cite{Lareau2000Social}, yet their simultaneous analysis is still rare. The goal in this last section is to address our third scientific question to quantify correlations between socioeconomic status, demographic characters, and the dynamics of purchase patterns of people. To tackle this question we define a weekly purchase vector for each user $u$. We introduce the  $w_u^k(d_i)$ fraction of money spent by user $u$ on a PCG $k$ on the day $d_{i\in [0,6]}$ such as $\sum_{i\in [0,6]}{w_u^k(d_i)}=1$. We also define the global fraction of $w_u(d_i)$ money spent by user $u$ during the weekday $d_{i\in [0,6]}$.

To quantify the relevance of demographic features and social status in the dynamics of purchase habits we consider a set of egos $X$. This set may represent a social class $s_{j\in[1-9]}$ (as defined in Section~\ref{sec:sociomeas}), or a set of people belonging to a given age group $a_{j\in [0-10]}$ (5-year brackets from $15-70$), or a group of egos with the same gender $g_{j\in [0-1]}$ (0 for female and 1 for male). For a given group $X$, whether we take a single PCG from $k\in K_{17}$, or a $\kappa \subset K_{17}$ set of PCGs, and measure
\begin{equation}
w^k_{X}(d_i)=\langle w^k_{u\in{X}}(d_i) \rangle \text{, } w^\kappa_{X}(d_i)=\langle w^\kappa_{u\in{X}}(d_i) \rangle \text{ and } w_{X}(d_i)=\langle w_{u\in{X}}(d_i)\rangle,
\label{eq:wukd}
\end{equation}
i.e. the $w^k_{X}(d_i)$ average weekly purchase vector of category $k$, the $w^\kappa_{X}(d_i)$ average weekly purchase vector of categories $\kappa$, and its global equivalent $w_{X}(d_i)$ characterising the overall spending pattern of class $X$.

\begin{figure*}[ht!]
\centering
\includegraphics[width=1.0\textwidth,angle=0]{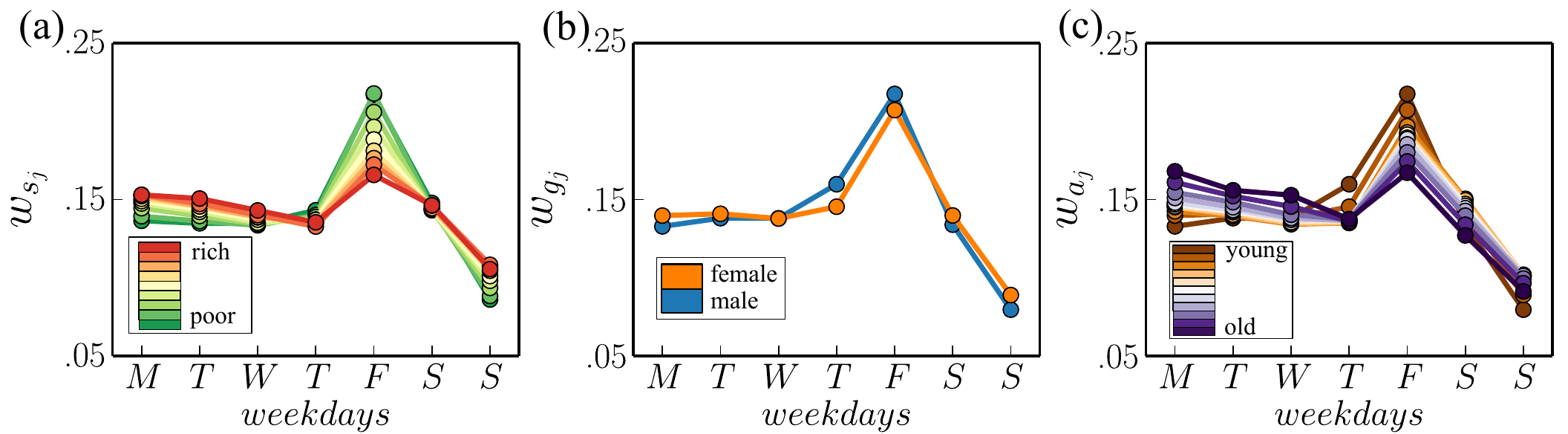}
\caption{\textbf{Weekly purchasing patterns according to social status, age and gender} As defined in Eq.\ref{eq:wukd}, \textbf{(a)} $w_{s_j}$ is the normalised distribution of spending in each day for each social classes \textbf{(b)}  $w_{g_j}$ is the normalised distribution of spending in each day for female $g_0$ and male $g_1$ \textbf{(c)} $w_{a_j}$ is the normalised distribution of spending in each day for each group of age. Nodes of the same distributions are connected by lines to lead the eye for better interpretation.
\label{fig:6}}
\end{figure*}

In order to depict dynamical purchase patterns, in Fig.~\ref{fig:6}a-c, we show the averaged weekly purchase vectors of groups of people with the same (a) socioeconomic status, (b) gender, and (c) age. There a global characteristic shape of the distributions appear over the weekdays, with a peak on Friday and a weaker signal on Sunday. One explanation of this overall pattern is rooted in the salary payment habits in the country, where salaries are typically payed weekly or bi-weekly and most of the time on Fridays, while on Sunday some shops are closed.

At first, in Fig.~\ref{fig:6}a we show the average weekly purchase vector calculated for each socioeconomic class. It shows a smooth substantial variation from the distribution of the poorest group (green) to the richest one (red) and suggests dependencies of weekly purchase activities on socioeconomic status. It varies continuously from the rich class, which spends rather constantly during the week but more on Monday, Tuesday, Wednesday and Sunday, whereas poorer egos spend more on Friday. Spending on Fridays varies from $w_{s_1}(4)=21.7\%$ to $w_{s_9}(4)=16.5\%$ going from the poorest to the richest classes. In Fig.~\ref{fig:6}b we show similar curves but now decoupled by gender. Interestingly, gender effects appear to be negligible here, assigning slightly higher activity on Monday and Saturday for female, and on Thursday and Friday for male. On the contrary, age seems to play a determinant role in purchase activity patterns as we see in Fig.~\ref{fig:6}c. Youngest (and at the same time poorer) people between age $20-25$, have a marked peak on Friday and spend slightly more on Thursday than other age groups. Contrary, the oldest people (60-65) get a higher activity in the beginning of the week (Monday, Tuesday and Wednesday) whereas 40-45 years-old people spend more than other group on Saturday and Sunday. These results suggest that the socioeconomic status and age of people are the most determinant characters when it turns to spending activity patterns, while contrary to the intuition, gender plays a less important role here.

\begin{figure*}[ht!]
\centering
\includegraphics[width=1.0\textwidth,angle=0]{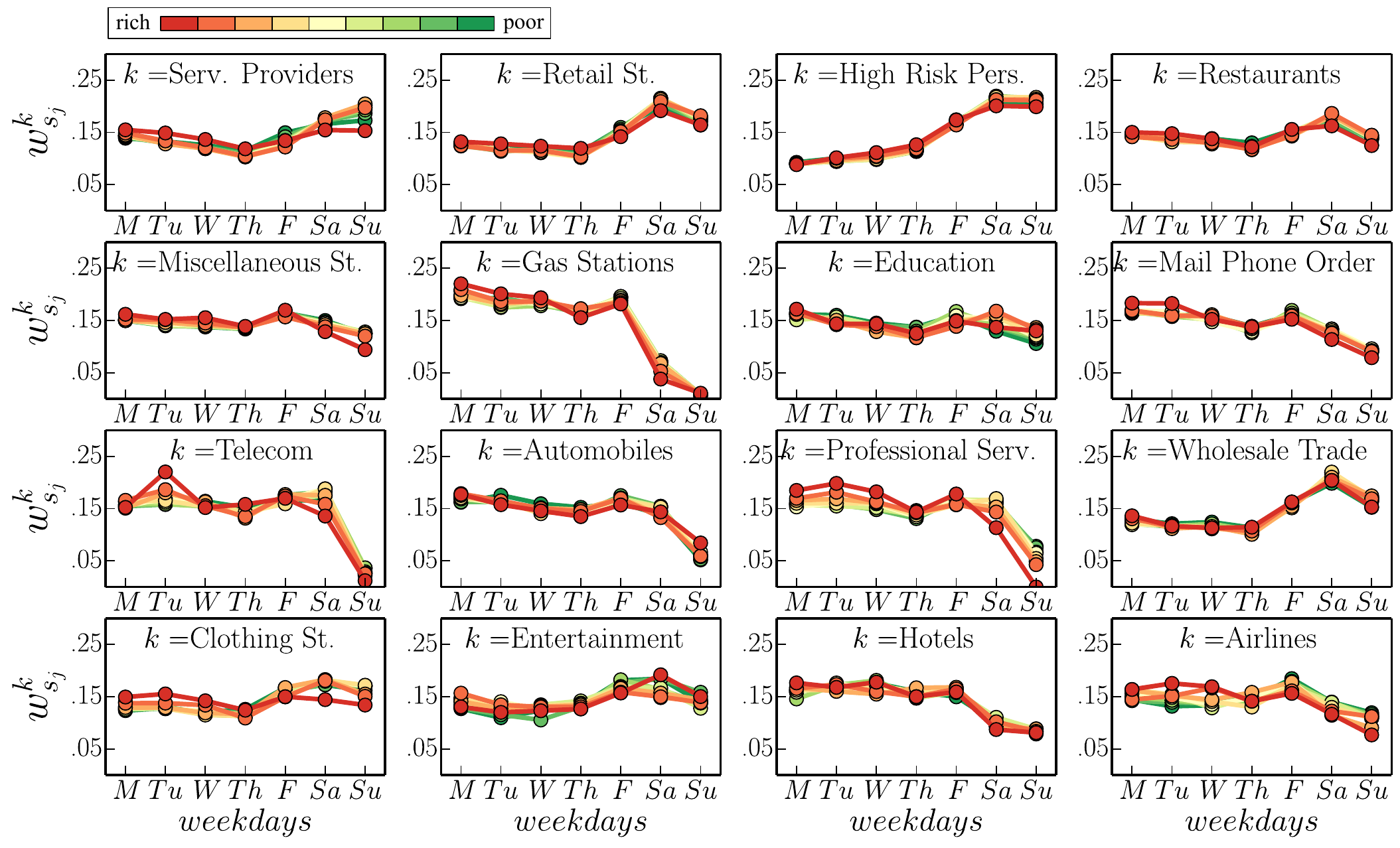}
\caption{\textbf{Weekly purchasing patterns according to social status for each PCG} As defined in Eq.\ref{eq:wukd}, $w^k_{s_j}$ is the normalised distribution of spending in each day for each social classes and in each PCGs $k\in K_{1-16}$. Nodes of the same distributions are connected by lines to lead the eye for better interpretation.
\label{fig:8}}
\end{figure*}

Going further, purchase activities during the week may also differ from one PCGs to another. To precisely show this, in Fig.~\ref{fig:8}, we depict the $w^k_{s_j}$ normalised distribution of spending in each day for each social class and in each PCGs $k\in K_{1-16}$. Some categories appear with the same temporal pattern like  \emph{Service Providers}, \emph{Retail Stores}, \emph{High Risk Personal Retail}, \emph{Wholesale Trade}, \emph{Restaurants}, \emph{Clothing St.} and \emph{Entertainment}, which all present relatively low activity before pay day and high activity in the end of the week (Friday, Saturday and Sunday). On the contrary, the PCGs like \emph{Gas Stations}, \emph{Automobiles}, \emph{Professional Services}, \emph{Hotels and Motels} and \emph{Airlines} appear with high activity during the working period (Monday to Thursday) but with lower values during the weekends. Even larger differences appear while comparing social group activities. The poorest class spends the most on Fridays for many PCGs such as \emph{Service Providers},  \emph{Gas Stations}, \emph{Education} , \emph{Automobiles}, \emph{Entertainment}, and \emph{Airlines}. On the contrary, richer people tend to spend on Saturdays on \emph{Entertainment}, \emph{Restaurants}, \emph{Wholesale Trade} and \emph{Education}. While PCGs such as \emph{Entertainment} and \emph{Education} have marked variation over social classes, others categories such as \emph{High Risk Personal Retail} and \emph{Wholesale Trades} appear to be closely the same for each socioeconomic group.

\begin{figure*}[ht!]
\centering
\includegraphics[width=0.65\textwidth,angle=0]{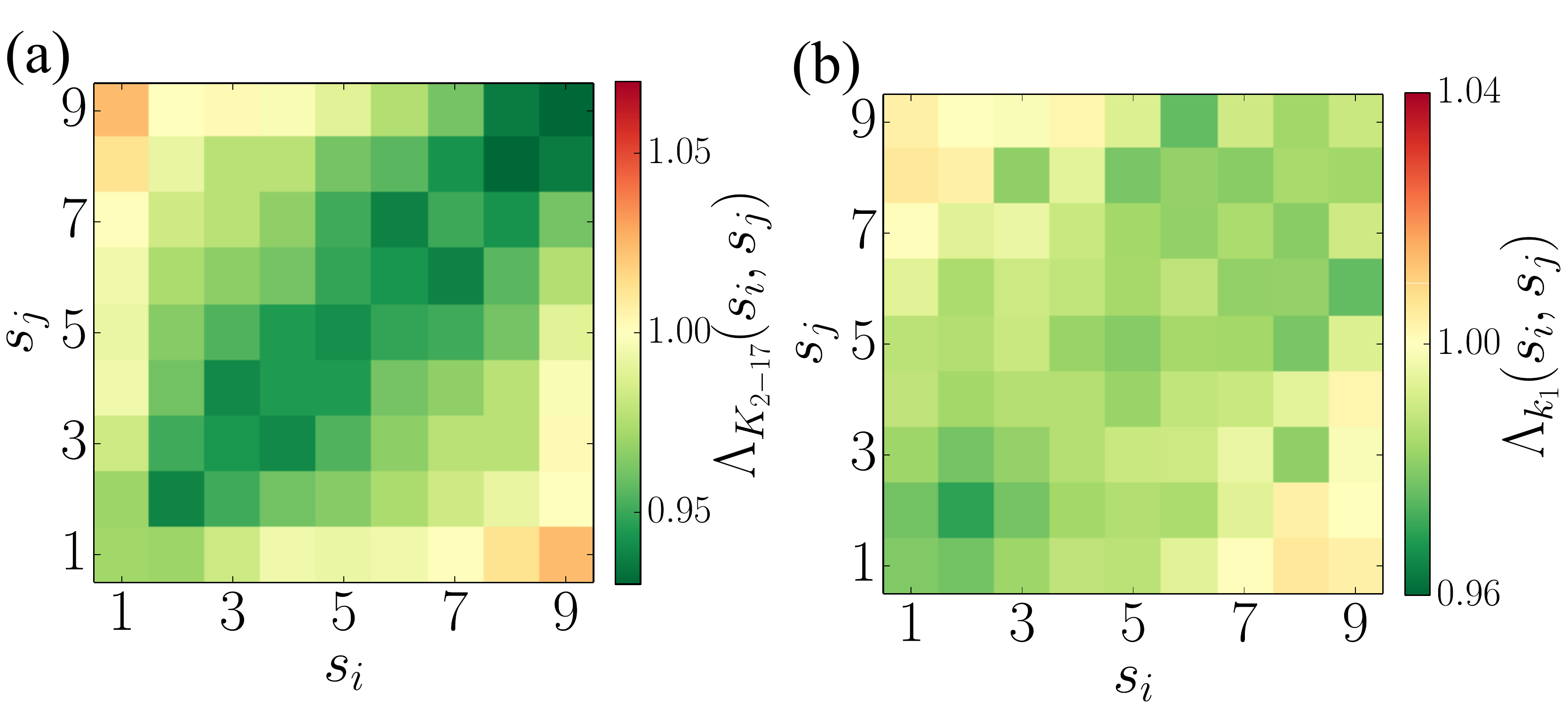}
\caption{\textbf{Dynamic purchase pattern in a socioeconomic network} \textbf{(a)} Heat-map matrix representation of the average $\Lambda_{K_{2-17}}(s_i,s_j)$ distance of the weekly purchase activity between pairs of socioeconomic classes. \textbf{(b)} Heat-map matrix representation of the average $\Lambda_{k_1}(s_i,s_j)$ distance of the weekly cash activity between pairs of socioeconomic classes.
\label{fig:9}}
\end{figure*}

As we studied in Section~\ref{sec:purchsocio}, network effects may increase behavioral similarities. We address this question by analysing an overall measure, which captures differences between the purchase dynamics of connected individuals positioned in the same or different socioeconomic classes. More precisely, we consider each social tie $(u,v)\in E$ connecting individuals $u\in s_i$ and $v\in s_j$, and we calculate the average Euclidean distance of their weekly purchase activity for a set $\kappa \subset K_{17}$ as
\begin{equation}
\delta_\kappa(s_i,s_j)=\langle \lVert w^\kappa_u(d_i)-w^\kappa_v(d_i) \rVert_2 \rangle_{u\in s_i, v\in s_j}
\end{equation}
Similar to our analysis in Section~\ref{sec:purchsocio}, we generate as a reference system a corresponding configuration network by taking randomly selected edge pairs from the underlying social structure and swap them without allowing multiple links and self loops and with the same parameters as earlier. In order to study the effects of the social network here we introduce the ratio $\Lambda_{K_{2-17}}(s_i,s_j)$ for each category group $k\in K_{2\text{-}17}$ and separately $\Lambda_{k_1}(s_i,s_j)$ for the category group $k_1$ (Service Providers) as:
\begin{equation}
\Lambda_{K_{2-17}}(s_i,s_j)=\frac{\delta_{K_{2-17}}(s_i,s_j)}{\langle \delta_{K_{2-17}}^{rn}\rangle (s_i,s_j)}  \hspace{.2in} \text{and} \hspace{.2in} \Lambda_{k_1}(s_i,s_j)=\frac{\delta_{k_1}(s_i,s_j)}{\langle \delta_{k_1}^{rn}\rangle (s_i,s_j)}
\end{equation}

Just as in case of purchase amount distributions in Fig.\ref{fig:3}, the normalised correlation matrix $\Lambda_{K_{2-17}}(s_i,s_j)$ in Fig.\ref{fig:9}a appears with a diagonal component. It indicates that the dynamics of purchase patterns of individuals connected in the original structure are actually more similar than expected in an uncorrelated network. On the other hand people from remote socioeconomic classes appear to be less similar than one would expect from the reference case. Note that even we found similar correlations between social ties and weekly cash patterns (in Fig.\ref{fig:9}b), these correlations appear to be somewhat weaker in this case.

\section{Conclusion}
\label{sec:concl}

In this paper we analysed a multi-modal dataset collecting the mobile phone communications and bank transactions of a large number of individuals from a Latin American country. This corpus allowed for an innovative global analysis both in terms of social network and in relation to the economical status and dynamical merchant patterns of individuals. We introduced a way to estimate the socioeconomic status of each individual together with several measures to characterise their purchasing habits. Using these information we identified distinct socioeconomic classes, which reflected strongly imbalanced distribution of purchasing power in the population. After mapping the social network of egos from mobile phone interactions, we showed that \textit{typical consumption patterns are strongly correlated with the socioeconomic classes and the social network behind}. We observed simultaneously these correlations on the individual and social class level.

In the second part of our study we detected correlations between co-purchased merchant categories and introduced a correlation network which in turn emerged with communities grouping homogeneous sets of categories together. We further analysed some multivariate relations between merchant categories and average demographic and socioeconomic features, and found \textit{meaningful patterns of correlations of co-purchased merchant categories.}

Finally we analysed dynamical purchase patterns and showed that while \textit{age and socioeconomic status are determinant factors}, gender seems to have a weaker role in differentiating between purchase habits. Similar to purchase distributions we found \textit{strong correlations in purchase dynamics between connected people belonging to the same socioeconomic class.}

Although our study is based on a combined dataset it comes with certain limitations. First of all we have access to the customers of a single mobile operator of the country. This sets some limitations as we cannot map out the complete social structure of the country, only receive a good approximation of it by considering communication links between company and non-company users. We are also limited by the bank dataset which provides us several informations about the purchasing habits of people, but its costumer set is only partially overlapping with the set of mobile costumers. On the other hand after careful selection of bank users with available social network informations we still obtained a set of $\sim 1$ million people, which allowed us to perform a meaningful statistical analysis about their consumer habits.

We can foresee several new directions to explore in the future. Possible tracks would be to better understand the role of the social structure and interpersonal influence on individual purchasing habits, the detection of causal correlated patterns of purchases, or to study how dynamics in communication and in purchases are intertwined, while the exploration of correlated patterns between commonly purchased brands may assign another promising direction. Beyond our general goal to better understand the relation between social and consuming behaviour, these results may enhance applications in several areas. Examples are marketing and advertising, where the knowledge of the reported correlations about co-purchased products, the effects of social ties, or the dynamical variance of purchase habits provide valuable information for the design of more efficient strategies. Other area of application is related to consumer behaviour prediction in terms of purchase and social influence, which is a rapidly growing field in the machine learning community. These are just a few examples where our results can be applied meaningfully. We hope that this contribution will foster further scientific studies on purchase-social behaviour and meaningful applications in several domains.

\vspace{.5in}

\textbf{Acknowledgements:}
We acknowledge the support from the SticAmSud UCOOL project, INRIA, and the SoSweet (ANR-15-CE38-0011-01) and CODDDE (ANR-13-CORD-0017-01) ANR projects. We thank M. Fixman for technical assistance.

\vspace{.1in}

This is a post-peer-review, pre-copyedit version of an article published in Social Network Analysis and Mining (Springer). The final authenticated version is available online.

\pagebreak

\begin{table*}[ht!]
\fontsize{6}{6.5}\selectfont
\centering
  \begin{tabular}{ | @{\hspace{1pt}} l @{\hspace{2pt}} | @{\hspace{1pt}} l @{\hspace{2pt}} | @{\hspace{1pt}} l @{\hspace{2pt}} | @{\hspace{1pt}} l @{\hspace{2pt}} |}
    \hline
742: Veterinary Serv. &	5300: Wholesale &	5950: Glassware, Crystal St. &	7523: Parking Lots, Garages  \\ \hline
763: Agricultural Cooperative &	5309: Duty Free St. &	5960: Dir Mark - Insurance &	7531: Auto Body Repair Sh.  \\ \hline
780: Landscaping Serv. &	5310: Discount Stores &	5962: Direct Marketing - Travel &	7534: Tire Retreading \& Repair  \\ \hline
1520: General Contr. &	5311: Dep. St. &	5963: Door-To-Door Sales &	7535: Auto Paint Sh.  \\ \hline
1711: Heating, Plumbing &	5331: Variety Stores &	5964: Dir. Mark. Catalog &	7538: Auto Service Shops  \\ \hline
1731: Electrical Contr. &	5399: General Merch. &	5965: Dir. Mark. Retail Merchant &	7542: Car Washes  \\ \hline
1740: Masonry \& Stonework &	5411: Supermarkets &	5966: Dir Mark - TV &	7549: Towing Serv.  \\ \hline
1750: Carpentry Contr. &	5422: Meat Prov. &	5967: Dir. Mark. &	7622: Electronics Repair Sh.  \\ \hline
1761: Sheet Metal &	5441: Candy St. &	5968: Dir. Mark. Subscription &	7623: Refrigeration Repair  \\ \hline
1771: Concrete Work Contr. &	5451: Dairy Products St. &	5969: Dir. Mark. Other &	7629: Small Appliance Repair  \\ \hline
1799: Special Trade Contr. &	5462: Bakeries &	5970: Artist's Supp. &	7631: Watch/Jewelry Repair  \\ \hline
2741: Publishing and Printing &	5499: Food St. &	5971: Art Dealers \& Galleries &	7641: Furniture Repair  \\ \hline
2791: Typesetting Serv. &	5511: Cars Sales &	5972: Stamp and Coin St. &	7692: Welding Repair  \\ \hline
2842: Specialty Cleaning &	5521: Car Repairs Sales &	5973: Religious St. &	7699: Repair Sh.  \\ \hline
4011: Railroads &	5531: Auto and Home Supp. St. &	5975: Hearing Aids &	7829: Picture/Video Production  \\ \hline
4111: Ferries &	5532: Auto St. &	5976: Orthopedic Goods &	7832: Cinema  \\ \hline
4112: Passenger Railways &	5533: Auto Access. &	5977: Cosmetic St. &	7841: Video Tape Rental St.  \\ \hline
4119: Ambulance Serv. &	5541: Gas Stations &	5978: Typewriter St. &	7911: Dance Hall \& Studios  \\ \hline
4121: Taxicabs &	5542: Automated Fuel Dispensers &	5983: Fuel Dealers (Non Auto) &	7922: Theater Ticket  \\ \hline
4131: Bus Lines &	5551: Boat Dealers &	5992: Florists &	7929: Bands, Orchestras  \\ \hline
4214: Motor Freight Carriers &	5561: Motorcycle Sh. &	5993: Cigar St. &	7932: Billiard/Pool  \\ \hline
4215: Courier Serv. &	5571: Motorcycle Sh. &	5994: Newsstands &	7933: Bowling  \\ \hline
4225: Public Storage &	5592: Motor Homes Dealers &	5995: Pet Sh. &	7941: Sports Clubs  \\ \hline
4411: Cruise Lines &	5598: Snowmobile Dealers &	5996: Swimming Pools Sales &	7991: Tourist Attractions  \\ \hline
4457: Boat Rentals and Leases &	5599: Auto Dealers &	5997: Electric Razor St. &	7992: Golf Courses  \\ \hline
4468: Marinas Serv. and Supp. &	5611: Men Cloth. St. &	5998: Tent and Awning Sh. &	7993: Video Game Supp.  \\ \hline
4511: Airlines &	5621: Wom Cloth. St. &	5999: Specialty Retail &	7994: Video Game Arcades  \\ \hline
4582: Airports, Flying Fields &	5631: Women's Accessory Sh. &	6211: Security Brokers &	7995: Gambling  \\ \hline
4722: Travel Agencies &	5641: Children's Wear St. &	6300: Insurance &	7996: Amusement Parks  \\ \hline
4784: Tolls/Bridge Fees &	5651: Family Cloth. St. &	7011: Hotels &	7997: Country Clubs  \\ \hline
4789: Transportation Serv. &	5655: Sports \& Riding St. &	7012: Timeshares &	7998: Aquariums  \\ \hline
4812: Phone St. &	5661: Shoe St. &	7032: Sporting Camps &	7999: Recreation Serv.  \\ \hline
4814: Telecom. &	5681: Furriers Sh. &	7033: Trailer Parks, Camps &	8011: Doctors  \\ \hline
4816: Comp. Net. Serv. &	5691: Cloth. Stores &	7210: Laundry, Cleaning Serv. &	8021: Dentists, Orthodontists  \\ \hline
4821: Telegraph Serv. &	5697: Tailors &	7211: Laundries &	8031: Osteopaths  \\ \hline
4899: Techno St. &	5698: Wig and Toupee St. &	7216: Dry Cleaners &	8041: Chiropractors  \\ \hline
4900: Utilities &	5699: Apparel Accessory Sh. &	7217: Upholstery Cleaning &	8042: Optometrists  \\ \hline
5013: Motor Vehicle Supp. &	5712: Furniture &	7221: Photographic Studios &	8043: Opticians  \\ \hline
5021: Commercial Furniture &	5713: Floor Covering St. &	7230: Beauty Sh. &	8049: Chiropodists, Podiatrists  \\ \hline
5039: Constr. Materials &	5714: Window Covering St. &	7251: Shoe Repair/Hat Cleaning &	8050: Nursing/Personal Care  \\ \hline
5044: Photographic Equip. &	5718: Fire Accessories St. &	7261: Funeral Serv. &	8062: Hospitals  \\ \hline
5045: Computer St. &	5719: Home Furnishing St. &	7273: Dating/Escort Serv. &	8071: Medical Labs  \\ \hline
5046: Commercial Equipment &	5722: House St. &	7276: Tax Preparation Serv. &	8099: Medical Services  \\ \hline
5047: Medical Equipment &	5732: Elec. St. &	7277: Counseling Services &	8111: Legal Services, Attorneys  \\ \hline
5051: Metal Service Centers &	5733: Music Intsruments St. &	7278: Buying/Shopping Serv. &	8211: Elem. Schools  \\ \hline
5065: Electrical St. &	5734: Comp.Soft. St. &	7296: Clothing Rental &	8220: Colleges Univ.  \\ \hline
5072: Hardware Supp. &	5735: Record Stores &	7297: Massage Parlors &	8241: Correspondence Schools  \\ \hline
5074: Plumbing, Heating Equip. &	5811: Caterers &	7298: Health and Beauty Spas &	8244: Business Schools  \\ \hline
5085: Industrial Supplies &	5812: Restaurants &	7299: General Serv. &	8249: Training Schools  \\ \hline
5094: Precious Objects/Stones &	5813: Drinking Pl. &	7311: Advertising Serv. &	8299: Educational Serv.  \\ \hline
5099: Durable Goods  &	5814: Fast Foods &	7321: Credit Reporting Agencies &	8351: Child Care Serv.  \\ \hline
5111: Printing, Office Supp. &	5912: Drug St. &	7333: Graphic Design &	8398: Donation  \\ \hline
5122: Drug Proprietaries &	5921: Alcohol St. &	7338: Quick Copy &	8641: Associations  \\ \hline
5131: Notions Goods &	5931: Secondhand Stores &	7339: Secretarial Support Serv. &	8651: Political Org.  \\ \hline
5137: Uniforms Clothing &	5932: Antique Sh. &	7342: Exterminating Services &	8661: Religious Orga.  \\ \hline
5139: Commercial Footwear &	5933: Pawn Shops &	7349: Cleaning and Maintenance &	8675: Automobile Associations  \\ \hline
5169: Chemicals Products  &	5935: Wrecking Yards &	7361: Employment Agencies &	8699: Membership Org.  \\ \hline
5172: Petroleum Products &	5937: Antique Reproductions &	7372: Computer Programming &	8734: Testing Lab.  \\ \hline
5192: Newspapers &	5940: Bicycle Sh. &	7375: Information Retrieval Serv. &	8911: Architectural Serv.  \\ \hline
5193: Nursery \& Flowers Supp. &	5941: Sporting St. &	7379: Computer Repair &	8931: Accounting Serv.  \\ \hline
5198: Paints &	5942: Book St. &	7392: Consulting, Public Relations &	8999: Professional Serv.  \\ \hline
5199: Nondurable Goods &	5943: Stationery St. &	7393: Detective Agencies &	9211: Courts of Law  \\ \hline
5200: Home Supply St. &	5944: Jewelry St. &	7394: Equipment Rental &	9222: Government Fees  \\ \hline
5211: Materials St. &	5945: Toy,-Game Sh. &	7395: Photo Developing &	9223: Bail and Bond Payments  \\ \hline
5231: Glass \& Paint St. &	5946: Camera and Photo St. &	7399: Business Serv. &	9311: Tax Payments  \\ \hline
5251: Hardware St. &	5947: Gift Sh. &	7512: Car Rental Agencies &	9399: Government Serv.  \\ \hline
5261: Nurseries \& Garden St. &	5948: Luggage \& Leather St. &	7513: Truck/Trailer Rentals &	9402: Postal Serv.  \\ \hline
5271: Mobile Home Dealers &	5949: Fabric St. &	7519: Mobile Home Rentals &  \\ \hline
  \end{tabular}
\caption{Codes and names of $271$ merchant categories used in our study. MCCs were taken from the Merchant Category Codes and Groups Directory published by American Express~\cite{MCCAmExp}. Abbreviations correspond to: Serv. - Services, Contr. - Contractors, Supp. - Supplies, St. - Stores, Equip. - Equipment, Merch. - Merchandise, Prov. - Provisioners, Pl. - Places, Sh. - Shops, Mark. - Marketing, Univ. - Universities, Org. - Organizations, Lab. - Laboratories.}
\label{table:mcc}
\end{table*}

\pagebreak


\end{document}